\documentstyle[12pt]{article}
\newcommand{\gt} {>}

\title{A DETERMINATION OF INTERFACE FREE ENERGIES}
\author{Alain Billoire \\
Service de Physique Th\'eorique de Saclay\thanks{
Laboratoire de la Direction des Sciences de la Mati\`ere du CEA.} \\
91191 Gif-sur-Yvette Cedex, France      \\
Thomas Neuhaus\\
Fakult\"at f\"ur Physik, Universit\"at Bielefeld \\
D-W 4800 Bielefeld, FRG\\
and Bernd A. Berg\thanks{On sabbatical from the Florida State University.} \\
Wissenschaftskolleg zu Berlin\\
Wallotstra\ss e 19, D-W 1000 Berlin 33, FRG}
\date{July 26 1993}
\begin{document}
\maketitle

We determine the interface free energy $F_{o.d.}$ between disordered
and ordered phases in the q=10 and q=20  2-d Potts models using the
results of multicanonical Monte Carlo simulations on $L^2$
lattices, and suitable finite
volume estimators. Our results, when extrapolated to the infinite volume
limit, agree to high precision with recent analytical calculations.
At the transition point $\beta_t$ the probability distribution function
of the energy exhibits two maxima. Their locations have $1/L^2$ corrections,
in contradiction with claims of $1/L$ behavior made in the literature.
Our data show a flat region inbetween the two maxima which characterizes
two domain configurations.
\vskip 3.5cm
\rightline {SPhT-93/065}
\vfill\eject

\section{Introduction}

  First order phase transitions play an important role in statistical
mechanics as well as in field theory, see for instance \cite{Juel}. For
example, the finite temperature phase transition of Quantum
Chromo Dynamics and the symmetry restoration phase transition
of the weak interaction theory are likely to be first order.
The value of the interface free energy is a
parameter of great importance. It determines the non--equilibrium
properties of the phase transition.
In this paper we address the question of how to determine
the value of the interface free energy
using Monte Carlo simulations.
For this purpose we consider much simpler models namely
the 2-d q=10 and q=20  Potts models \cite{Potts}, which in statistical
mechanics are prototype models of systems with
first order phase transitions.

  Recently two interesting analytical results have been obtained for
Potts models in the first order phase transition region.  One is a
rigourous theory of finite size scaling (FSS)
\cite{Borgs_deux,Borgs_Miracle}, which has been proven for
large values of q, although it is likely to hold for all $q>4$. However, it
has been found numerically that very large lattices are required in order
to observe the predicted behavior \cite{billneuber}.
The other result are calculations of the 2-d spin-spin correlation length
at the infinite volume phase transition point $\beta_t$ \cite{Zitta,BuffWall}.
The exact value of the order--disorder interface free energy $F_{o.d.}$
follows \cite{BoJa}. These results could then be compared with previously
obtained Monte Carlo (MC) results, of which the multicanonical \cite{BN,JBK}
proved to be most accurate. Table 1 displays the exact interface free
energy for q=7, 10 and 20 together with MC  estimates.
Although the overall accuracy of the MC estimates is quite satisfactory,
it seems that there are uncontrolled systematic errors larger
than the estimated statistical error. It is thus a challenge to improve the
methods employed for infinite volume estimates from finite volume
numerical calculations. These questions are addressed in the next two
sections of the paper.
In section 4 we analyze
multicanonical MC data for q=10 and 20. Section 5 summarizes the final
conclusions.

\section{Finite Volume Energy Distribution}

  We consider $L^d$ lattices with periodic boundary conditions. In the
2-d Potts model the energy density ranges in steps of $1/L^2$ in the
interval $ -2 \leq E \leq 0$.
Numerical simulations yield statistical estimates
of the canonical probability density $P_L(\beta,E)$. In this paper we are
concerned with the shape of $P_L(\beta,E)$ in the vicinity of the transition
point $\beta_t$, in order to extract the interface free energy $F_{o.d.}$.
In the transition region one expects for $P_L(\beta,E)$ the existence of two
distinct maxima $P_L^{max,o}$ and $P_L^{max,d}$, corresponding to the ordered
and disordered bulk states of the system. These are  centred at
separated values of the energy $E^{max,o}_L$ and $E^{max,d}_L$.  Inbetween
there will be a value of the energy $E^{min}_L$ where $P_L(\beta,E)$ takes its
minimal value $P_L^{min}$. States with energies close to $E^{min}_L$
correspond to configurations containing two interfaces.

Following \cite{Borgs_deux}, we can write the
partition function for the q-state Potts model on a $L^d$ lattice with
periodic boundary conditions in the form
\begin{equation}
Z_L(\beta)  =   e^{- L^d \beta f_d(\beta)} + q e^{- L^d \beta f_o(\beta)}
 +   {\cal O}(e^{-b L})  e^{- \beta f(\beta) L^d }
\hskip .8cm ;\  b > 0.
\label{partition}
\end{equation}
Here are $f_o(\beta)$ and $f_d(\beta)$ smooth $L$ independent
functions, representing the free energy densities of the
ordered and disordered phases. Furthermore one has
$f(\beta)=\min\{f_o(\beta),f_d(\beta)\}$.
At $\beta_t$ both free energies become equal to
$f(\beta_t)$.  In the vicinity of $\beta_t$ they possess the expansions
\begin{equation}
\beta f_i(\beta)=
\beta f(\beta_t)
+(\beta-\beta_t) E_i
+{\frac{(\beta-\beta_t)^2}{2}} ( {\frac{-C_i}{\beta_t^2}})
+ ...
\label{expans}
\end{equation}
with $i=o,d$ respectively. The  $C_i$ and the $E_i$ denote infinite volume
specific heat and energy densities at $\beta_t$.
In 2-d they have the exact values
\cite{Baxter} $E_o=-1.66425$, $E_d=-0.96820$, $C_d-C_o=0.44763$
for q=10 and $E_o=-1.82068$, $E_d=-0.62653$, $C_d-C_o=0.77139$ for q=20.
The large $q$ expansion of \cite{BLM} gives $C_o=18.06(4)$ and
$C_o=5.362(3)$,
in addition numerical simulations indicate $C_o=12-18$ for q=10 and
$C_o=5.2(2)$ for q=20 \cite{BillMoLa,billneuber}.

We obtain the density of states $\Omega_L(E)$ by inverse Laplace
transform.
\begin{equation}
\Omega_L(E)= {L^d\over 2 \pi i} \int^{c+i \infty}_{c-i \infty}
   e^{ L^d \zeta E} Z_L(\zeta) \  d\zeta  .
\end{equation}
We now consider the specific form $Z_L(\beta)=\exp(-L^d g(\beta))$, which is
the form of the partition function of each of the bulk phases of
eq.(\ref{partition}), neglecting terms ${\cal O}(e^{-b L})$
(we understand that this is not a valid procedure
for values of $E$ close to $E_L^{min}$, see below). It follows
\begin{equation}
\Omega_L(E)= {L^d\over 2 \pi i} \int^{c+i \infty}_{c-i \infty}
   e^{ L^d ( \zeta E - g(\zeta))} \ d\zeta .
\end{equation}
This integral is computed using the steepest descent method.  The  saddle
point is at $\zeta=\beta_s(E)$ which is solution of the equation
\begin{equation}
E={\partial g(\zeta)\over \partial\zeta}\Biggl\vert_{\zeta=\beta_s(E)}.
\end{equation}
$\beta_s(E)$  is the value of $\beta$ for
which $E$ is the internal energy of
the system. Since
\begin{equation}
{\displaystyle\partial^2 g(\zeta)\over
 \partial^2\zeta}\Biggl\vert_{\beta=\beta_s(E)}
=-T^2 C(\beta_s(E))
\end{equation}
is negative the straight path $\zeta=\beta_s+i x$, with $x$ real,
is the steepest descent, and we obtain the saddle point approximation
\begin{equation}
\Omega_L(E) \ \approx\ { L^{d/2}
e^{\displaystyle L^d (E \beta_s(E) -g(\beta_s(E) )}
\over \sqrt{-2 \pi {\displaystyle \partial^2 g(\zeta)
\over\displaystyle \partial^2\zeta}\Biggl\vert_{\zeta=\beta_s(E)} } } .
\label{new}
\end{equation}
The gaussian approximation for
$P_L(\beta,E) = {1\over Z_L(\beta)}e^{- \beta E L^d} \Omega_L (E)$
is obtained expanding $E$ around $\bar{E}_i(\beta)$, the
internal energy at $\beta$, then
\begin{equation}
P_L(\beta,E)  \approx \sum_{i=d,o} (\delta_{d,i}+q\delta_{o,i})
\sqrt{L^d\over{2  \pi  T^2 C_i(\beta)}} \
e^{-\displaystyle {{(E-{\bar E}_i(\beta))^2 L^d } \over{2 T^2 C_i(\beta)}}}\
e^{-\displaystyle\beta f_i(\beta) L^d} .
\end{equation}
This form corresponds to
Binder's ansatz for $P_L(\beta,E)$ in the vicinity of a first order
transition point \cite{Challa}. In this approximation the locations of
the maxima $E_L^{max,i}$ of $P_L(\beta,E)$ show no finite volume dependence.
In order to find the dependence, one keeps one more term in the expansion
of $g(\beta)$. Then $g^{''}|_{\zeta=\beta_S(E)}\approx g^{''}|_{\zeta=\beta}+
(E-{\bar E(\beta)}) \ {g^{'''}|_{\zeta=\beta}/ g^{''}|_{\zeta=\beta})}$,
 $E_L^{max,i}\approx{\bar E}_i+(g^{'''}/(2 g^{''} L^d)$.
This effect is due the asymmetry of the fluctuations inside each bulk phase,
and is not the one advocated in \cite{Lee}.

A few  remarks on the two Gaussian approximation are appropriate here.

$\bullet$ The two Gaussian approximation predicts at the transition point
$\beta_t$
a ratio $R_q=P_L^{max,o}/P_L^{max,d}=q \sqrt{C_d/C_o}$. Putting in
actual numbers for $C_o$ and $C_d$ we find $R_{10}=10.12$ for q=10 and
$R_{20}=21.4$ for q=20. The ratio of the weights of the ordered and
disordered states is precisely equal to $q$.
This follows from  eq.(\ref{partition})
and does not rely on the gaussian approximation.

$\bullet$ The widths of the two peaks of $P_L(\beta,E)$ are predicted, namely
\begin{equation}
\sigma_i= { \frac{1}{L^{\frac{d}{2}}\beta_t} }\sqrt{ 2 C_i } .
\label{width}
\end{equation}

$\bullet$  Inbetween the two maxima the Gaussian approximation is incorrect.
In this region mixed phase configurations dominate, this is the subject of
the next section.

\section{Mixed Phase Configurations}

For energies between the two peaks and large $L$, the probability to be
in a pure phase is suppressed like
$\exp [-(E-{\bar E})^2 L^d/ (2 T^2 C_i)]$.
Mixed phase configurations, where each phase occupies a macroscopic
fraction of the system, become dominant \cite{Binder} since they are
only suppressed
like $\exp (-\sigma_{o.d.}\ {\rm const.} L^{d-1})$, where $\sigma_{o.d.}$ is
the order disorder interface tension and ${\rm const.} L^{d-1}$ the
interface area (interface length for 2-d). Sufficiently close to
$E^{min}_L$ the configurations with minimum interface area
contains two planar interfaces, closed by periodic boundary conditions.
Closer to the peaks configurations with minimum interface are made of
a single macroscopic bubble of one phase inside the other phase.
The shape of the bubble is given by the Wulff construction, it is
spherical when the correlation length becomes infinite. The interface
area of a two planar interface configuration is $2 L^{d-1}$.
In the 2-d case, it dominates over the spherical bubble
configuration for

\begin{equation}
E^- < E < E^+
\end{equation}
with
\begin{eqnarray}
E^- = E_o - {\frac{1}{\pi}}(E_o-E_d) , \\
\label{bub1}
E^+ = E_d + {\frac{1}{\pi}}(E_o-E_d) .
\label{bub2}
\end{eqnarray}

Such a behavior has been rigourously shown to occur for the order
order magnetic transitions of the 2-d Ising model in the broken
phase \cite{Schlo}. Numerically it was observed for the 2-d and
3-d Ising models \cite{Ising}.
Two planar interface configurations, with a fraction $x$ of the volume in
the ordered phase, contribute to the partition function as

\begin{equation}
Z^{\rm mixed}(x) \propto e^{- x L^d \beta f_o(\beta)}
\times  e^{ - (1-x)L^d \beta f_d(\beta)} \times L^p e^{-F_{o.d.}L^{d-1}}  .
\label{Zstrip}
\end{equation}

The first two factors are contributions from the pure phases. The factor
$L^p \exp (-F_{o.d.} L^{d-1})$ comes from interface effects,
$F_{o.d.}=2\sigma_{o.d.}$ and
$p=d-1$ in the capillary wave approximation \cite{CO,Wiese}.

At the infinite volume transition point $\beta_t$ the free
energy densities $f_o$ and $f_d$  equal $f(\beta_t)$. $Z^{\rm mixed}(x)$
will then asymptotically have the $x$-independent form
\begin{equation}
Z^{\rm mixed}\mid_{\beta=\beta_t} \propto e^{- L^d \beta_t f(\beta_t)}
  \times L^p\ e^{-F_{o.d.}L^{d-1}}.
  \label{Zstrip2}
\end{equation}

It means that in the region around $E^{min}_L$ the probability distribution
$P_L(\beta_t,E) =
Z^{\rm mixed} / Z \propto L^p\ \exp (-F_{o.d.} L^{d-1})$ is flat.
This formula could be used to
determine the interface tension. However, we found it more convenient to
use the ratio $\sqrt{P_L^{max,o} P_L^{max,d}}/P_L^{min} \propto
L^{-p+d/2}  \exp (F_{o.d.} L^{d-1})$, and introduce the finite volume
estimator
\begin{eqnarray}
F(L) &=& {1\over 2 L^{d-1}} \ln \lbrack{P_L^{max,o}P_L^{max,d}
\over (P_L^{min})^2}\rbrack \mid _{\beta=\beta_t} \\
     &=& F_{o.d.} + {a_1 \over L^{d-1}} +
(-p+{d\over 2}) { \ln (L) \over L^{d-1}} + {a_2 \over L^{d}}
+ {a_3 \over L^{d+1}} + ...~.
\label{fitform}
\end{eqnarray}
It is convenient to consider $P_L(\beta_t,E)$ because one can
check whether it has a flat part as it should.
If it does not, the value of $F_{o.d.}$
extracted from eq. (\ref{fitform}) is questionable. However
this equation, with $L^d(\beta-\beta_t)$ dependent $a_i$'s,
holds asymptotically for all $\beta$ in a $\propto 1/L^d$ neighborhood of
$\beta_t$ (e.g. at the equal heights point like in \cite{BN}).

\section{Numerical Results}

   The simulations have been performed with the multicanonical
algorithm \cite{BN}, for more information we refer to \cite{billneuber}.
For q=10 we have simulated lattices in the range from $12^2$ to $100^2$,
for q=20 from $16^2$ to $70^2$. The statistics for q=10 is given in \cite{BN},
for q=20 in \cite{billneuber}, and additional q=20 statistics is collected
in Table 2.
Errors quoted in this paper are bias corrected jackknife estimates. The
simulations have been performed on the Saclay Cray X-MP, on RISC stations of
the SCRI workstation cluster and on an ALPHA station at Bielefeld University.

\begin{figure} [htbp]
\vfill\penalty -5000\vglue 14cm
\includegraphics{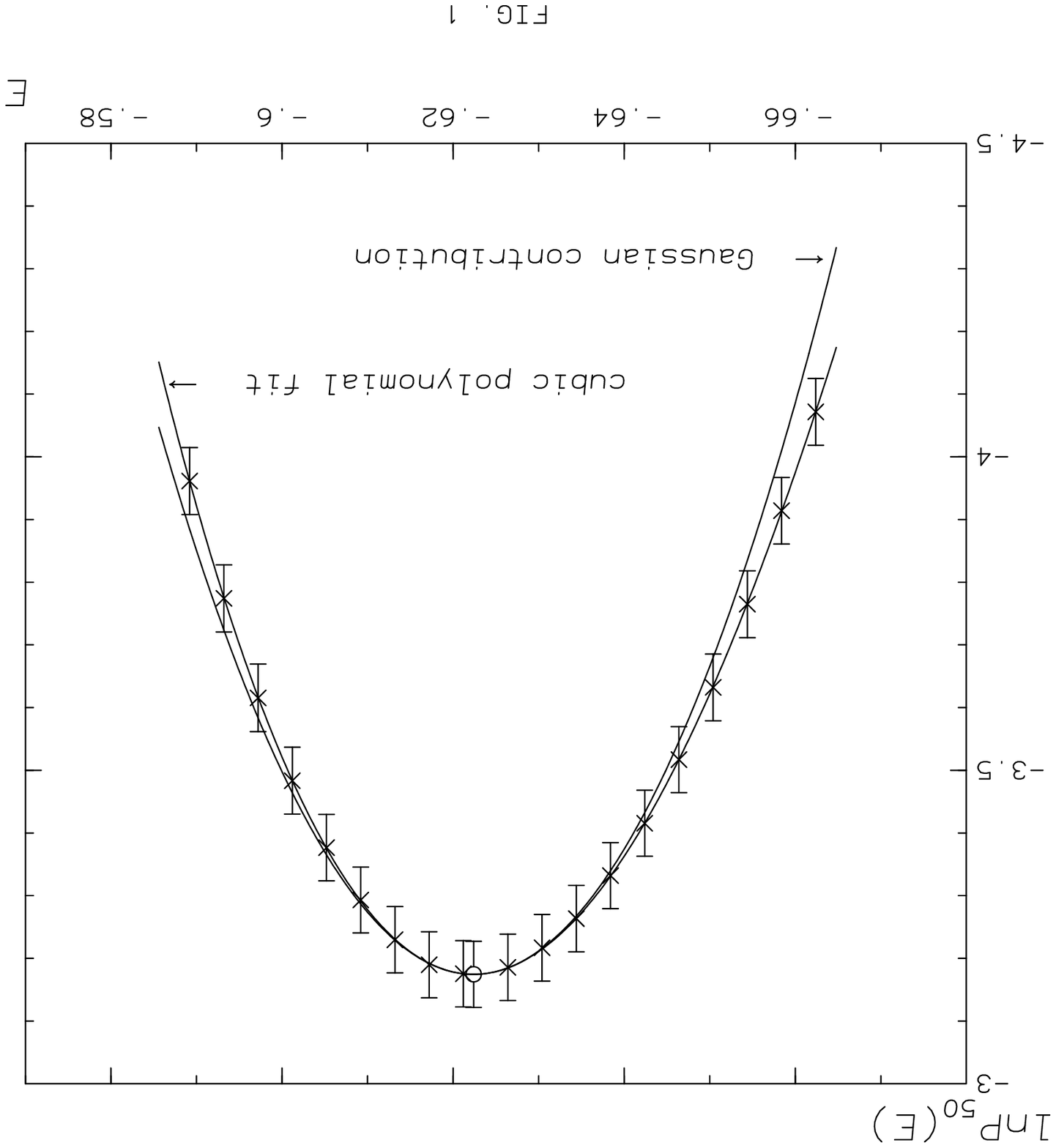}
\caption{ Plot of  $\ln~P_L(\beta_t,E)$ on a $50^2$ lattice  for the
q=20 Potts model in the vicinity of the disordered state. The crosses denote
(selected) data with their error bars. The circle denotes the maximum
value $P_L^{max,d}$ with error bar. The two curves display
a cubic polynomial fit to $\ln~P_L(\beta_t,E)$
and the Gaussian contribution to this fit.
}
\label{50maxd}
\end{figure}

\begin{figure} [htbp]
\vfill\penalty -5000\vglue 14cm
\includegraphics{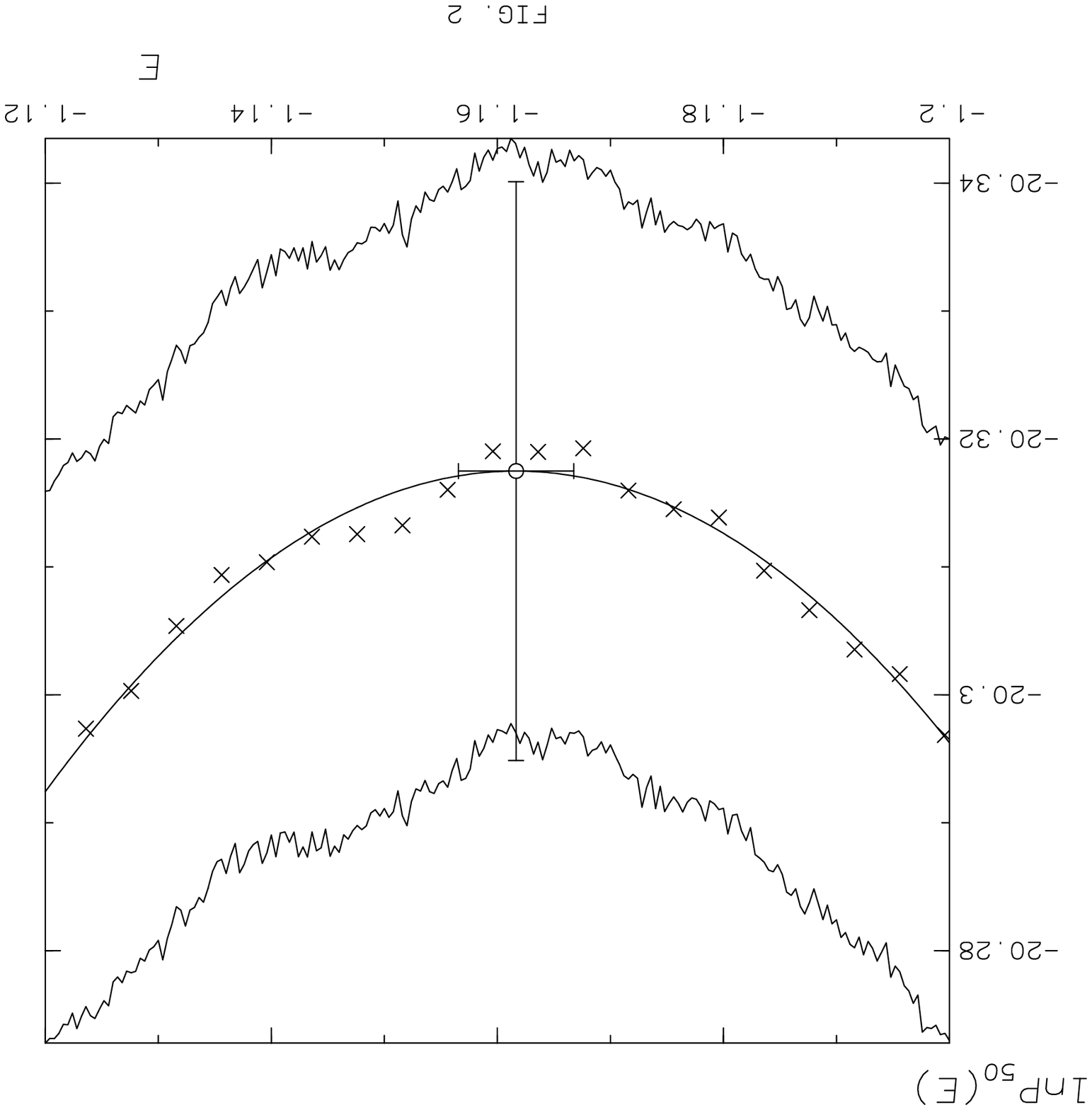}
\caption{Plot of $\ln~P_L(\beta_t,E)$ on a $50^2$ lattice in the q=20 Potts
model
in vicinity of $P_L^{min}$. The crosses denote selected
data. The wiggling curves indicate the error interval of the data, while
the circle with errror bars denotes our estimate of $P_L^{min}$
and $E_L^{min}$.
}
\label{50min}
\end{figure}

For the determination of the quantities $P_L^{max,o}$, $P_L^{max,d}$,
$E_L^{max,o}$ and $E_L^{max,d}$ we adopt the following procedure.
After locating approximate values we fit the probability function
$P_L(\beta_t,E)$ in the vicinity of the maxima on an energy interval
corresponding to a decrease of $P_L(\beta_t,E)$ by at most a factor
$1/e$ of its maximal value. We use the cubic form
\begin{equation}
\ln~P_L(\beta_t,E)= {\rm const} +  {(E-E_L^{max,i})^2 \over \sigma_i}
+ \alpha (E-E_L^{max,i})^3 .
\label{maxfit}
\end{equation}
We have checked that the fits gave acceptable $\chi^2/d.o.f$ values.
Figure \ref{50maxd} displays an example for the q=20 model on a $50^2$ lattice.
It can be seen that the probability distribution in vicinity of the
maximum shows a sizable asymmetry, demonstrating that the cubic term
is needed in order to give a good description of the data.
For the determination of the quantities $P_L^{min}$ and
$E_L^{min}$ we found it sufficient to use a quadratic fit analogue to
eq.(\ref{maxfit}) with $\alpha=0$. An example of such a fit is displayed
in Figure \ref{50min}.

\begin{figure} [htbp]
\vfill\penalty -5000\vglue 14cm
\includegraphics{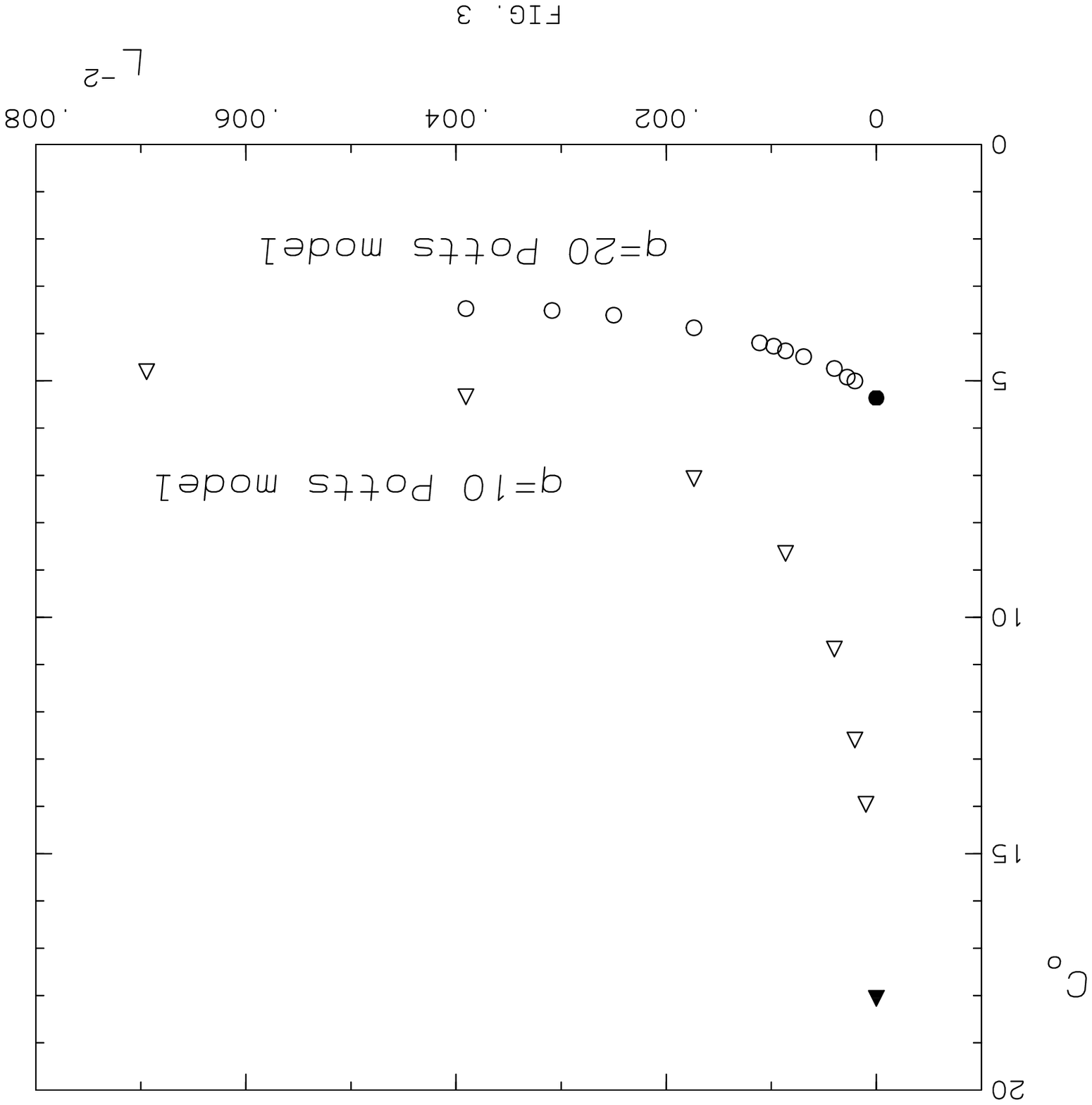}
\caption{   Estimates of $C_o$ for the q=10 (circles) and q=20 Potts
(triangles) models,
as extracted from the width of the ordered peak of $P_L(\beta_t,E)$.
The full symbols denote asymptotic results from the large
q-expansion.}
\label{cv}
\end{figure}

  Using eq.(\ref{width}) and the width parameters $\sigma_i$ for $i=o,d$
as obtained by the fit (\ref{maxfit}) we determine finite volume
estimators of the specific heats $C_o(L)$ and $C_d(L)$. The results are
collected in Table 3 for  q=10 and in Table 4 for q=20.
Figure \ref{cv} displays a plot of $C_o(L)$ for  q=10 and q=20,
together with
asymptotic values predicted by a large q-expansion \cite{BLM}.
In case of q=20 our data for $C_o(L)$ converge towards a value consistent
with the large $q$ expansion results and
the FSS study \cite{billneuber} that gives
$C_o  = 5.2 (2)$. In addition, the data for $C_d(L)-C_o(L)$ approach, as $L$
grows, the exactly known infinite volume limit.
For the q=10 model the data
overshoot the value $C_o = 12.7 (3)$, obtained from the FSS of the
extrema of the specific heat \cite{BillMoLa}. The value indicated by
the large q-expansions is $C_o = 18.06 (4)$.
A value $C_o \approx 18$ was obtained in \cite{BillMoLa} from
the FSS of the specific heat at $\beta_t$. Much larger lattices would be
required in order to see $C_o(L)$ approach this limit
(The correlation length \cite{BuffWall} at $\beta_t$ is
$\xi_{disorder}=10.56$ for $q=10$ and $2.70$ for $q=20$).

\begin{figure} [htbp]
\vfill\penalty -5000\vglue 14cm
\includegraphics{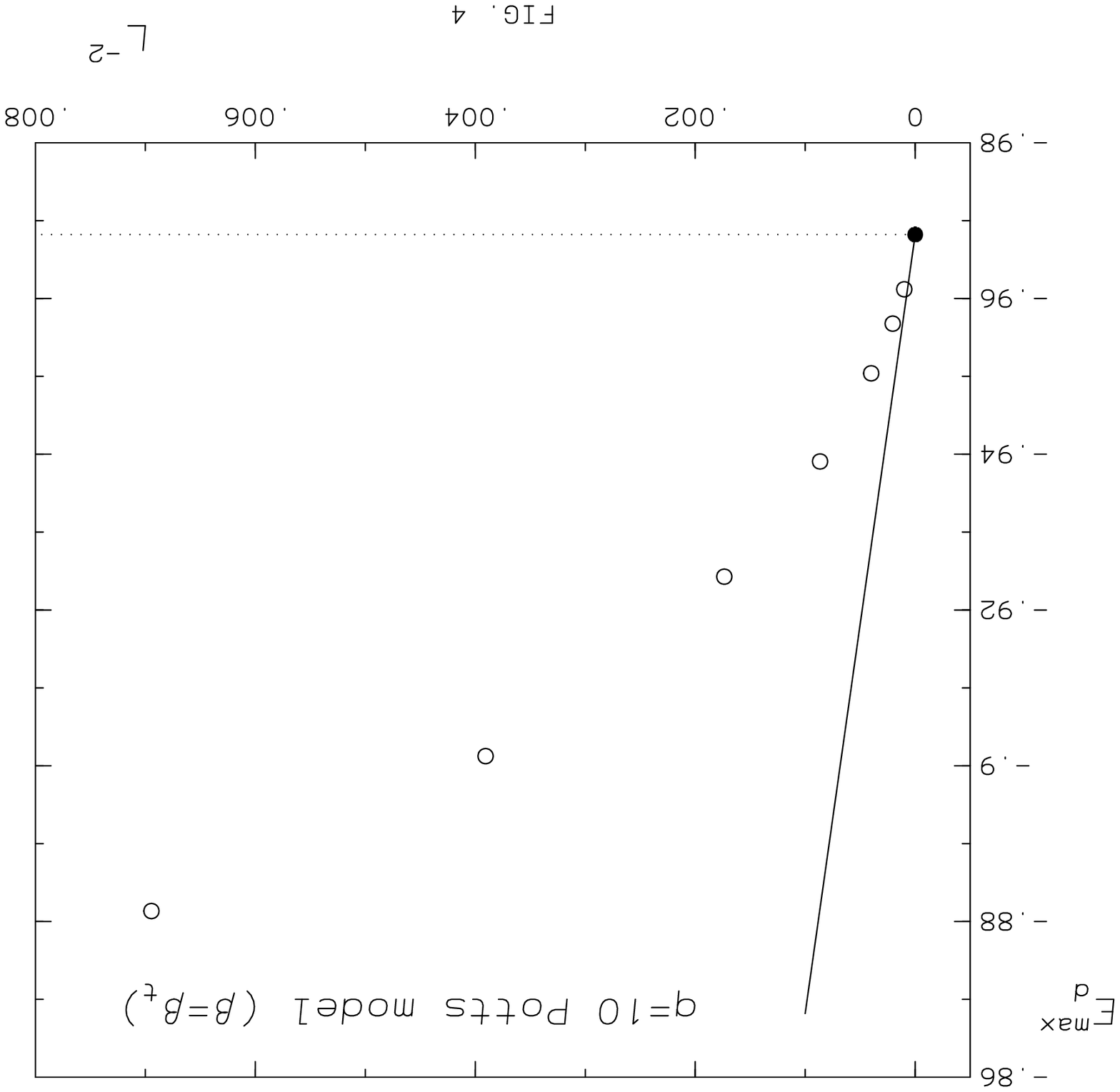}
\caption{   Locations of the maxima of disorderd states $E_L^{max,d}$
at $\beta_t$ in the q=10 Potts model as function of the inverse volume.
The full symbol and dashed line denote the infinite volume prediction.
The full line has a slope $g'''/(2g'')$ as predicted by the large
q expansion [12]. }
\label{Fig4}
\end{figure}

\begin{figure} [htbp]
\vfill\penalty -5000\vglue 14cm
\includegraphics{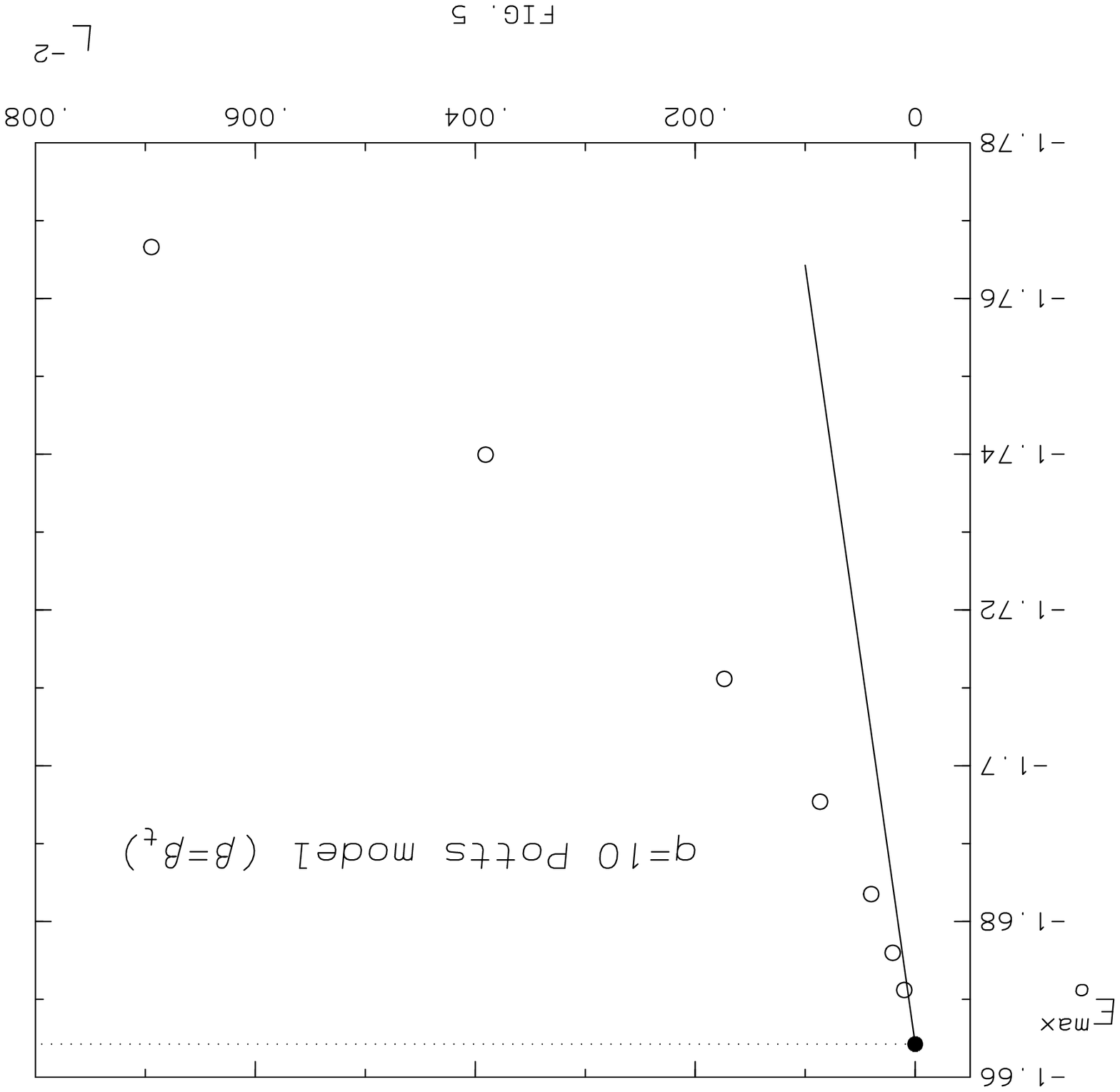}
\caption{   Locations of the maxima of ordered states $E_L^{max,o}$
at $\beta_t$ in the q=10 Potts model as function of the inverse volume.
The symbols and lines have the same meaning as in
Fig. 4. }
\label{Fig5}
\end{figure}

\begin{figure} [htbp]
\vfill\penalty -5000\vglue 14cm
\includegraphics{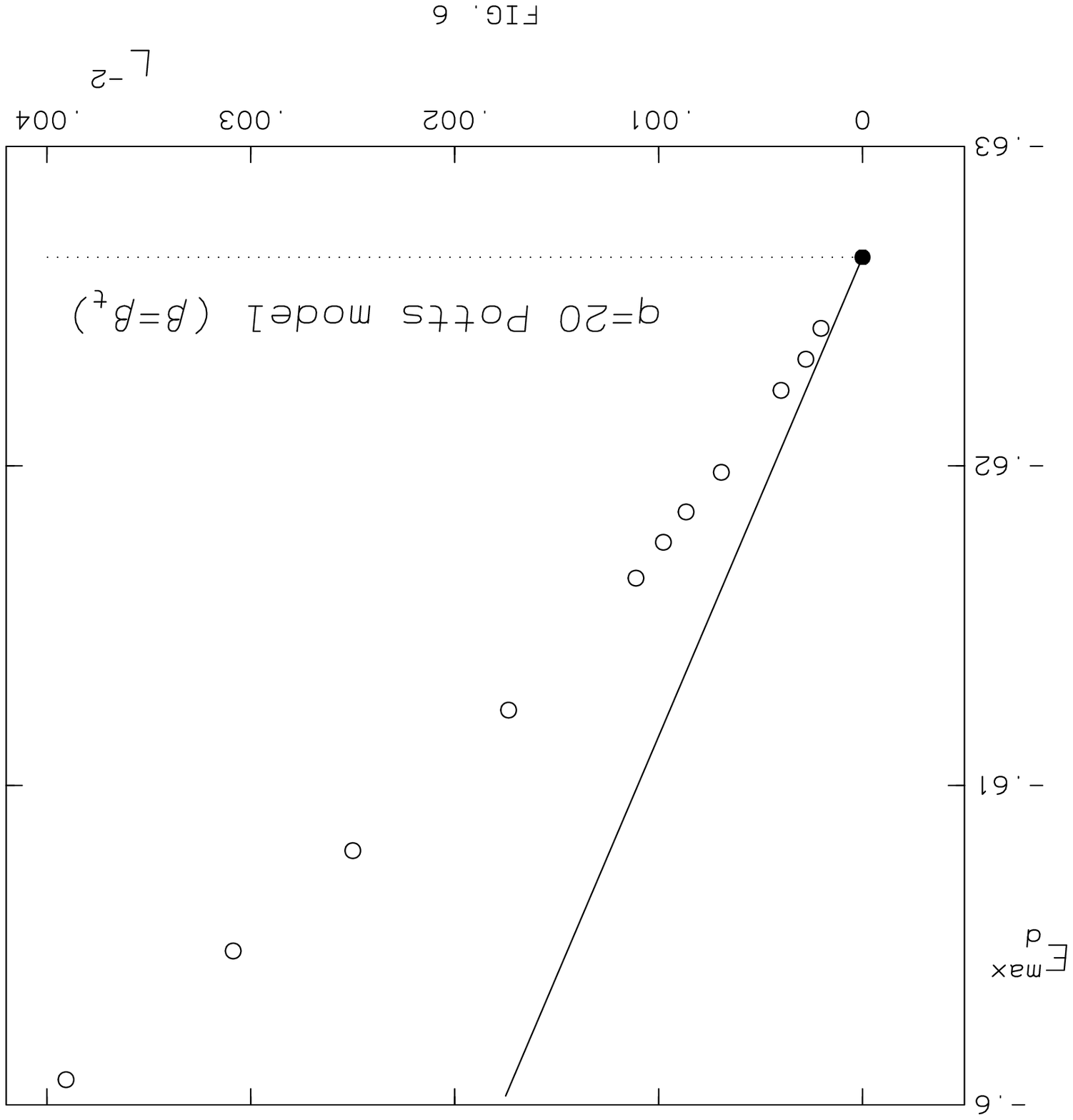}
\caption{   Locations of the maxima of disordered states $E_L^{max,d}$
at $\beta_t$ in the q=20 Potts model as function of the inverse volume.
The symbols and lines have the same meaning as in
Fig. 4. }
\label{Fig6}
\end{figure}

\begin{figure} [htbp]
\vfill\penalty -5000\vglue 14cm
\includegraphics{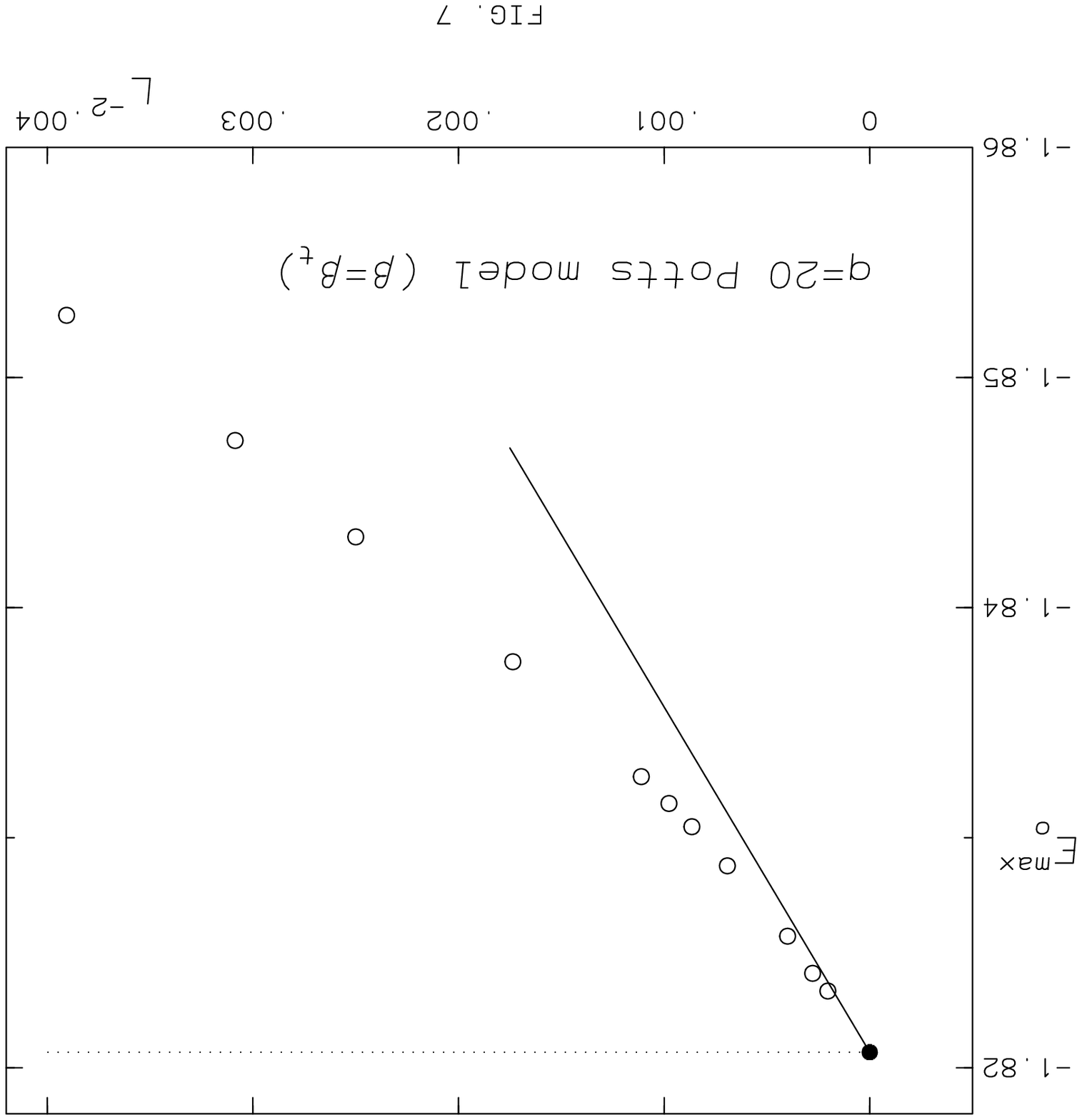}
\caption{   Locations of the maxima of ordered states $E_L^{max,o}$
at $\beta_t$ in the q=20 Potts model as function of the inverse volume.
The symbols and lines have the same meaning as in
Fig. 4. }
\label{Fig7}
\end{figure}

In figure \ref{Fig4} to \ref{Fig7} we display the locations
of the maxima $E_L^{max,o}$
and $E_L^{max,d}$ of the energy probability distribution function at
$\beta_t$ as function of the variable $1/L^d$, compared with the
prediction $E^{max,i}_L={\bar E}_i + const. /L^d$
with slopes \cite{BLM} 100 ($q=10$) and 15 ($q=20$).
The data are collected also
in Tables 3 and 4. For our largest lattices the predicted linear dependence
on $1/L^d$ is born out.
The deviations observed can be reproduced \cite{Morel}
within the formalism of \cite{BLM}. This rules out
the $1/L$
behavior predicted in \cite{Lee}.
For the larger sized lattices the ratio $R_q$ become consistent
with the theoretical prediction $q \sqrt{C_d/C_o}$.
These tables contain also our estimates for the partition functions ratios
$Z_L^o(\beta_t)/Z_L^d(\beta_t)$. They are defined by summing the
probability distribution
function at $\beta_t$ over corresponding energy ranges $E<E_L^{min}$,
resp. $E>E_L^{min}$. These ratios are consistent with the value q as
predicted by theory.  Unfortunately, due to small
numbers of tunneling events, they have  sizable
errors on our largest lattices.

\begin{figure} [htbp]
\vfill\penalty -5000\vglue 14cm
\includegraphics{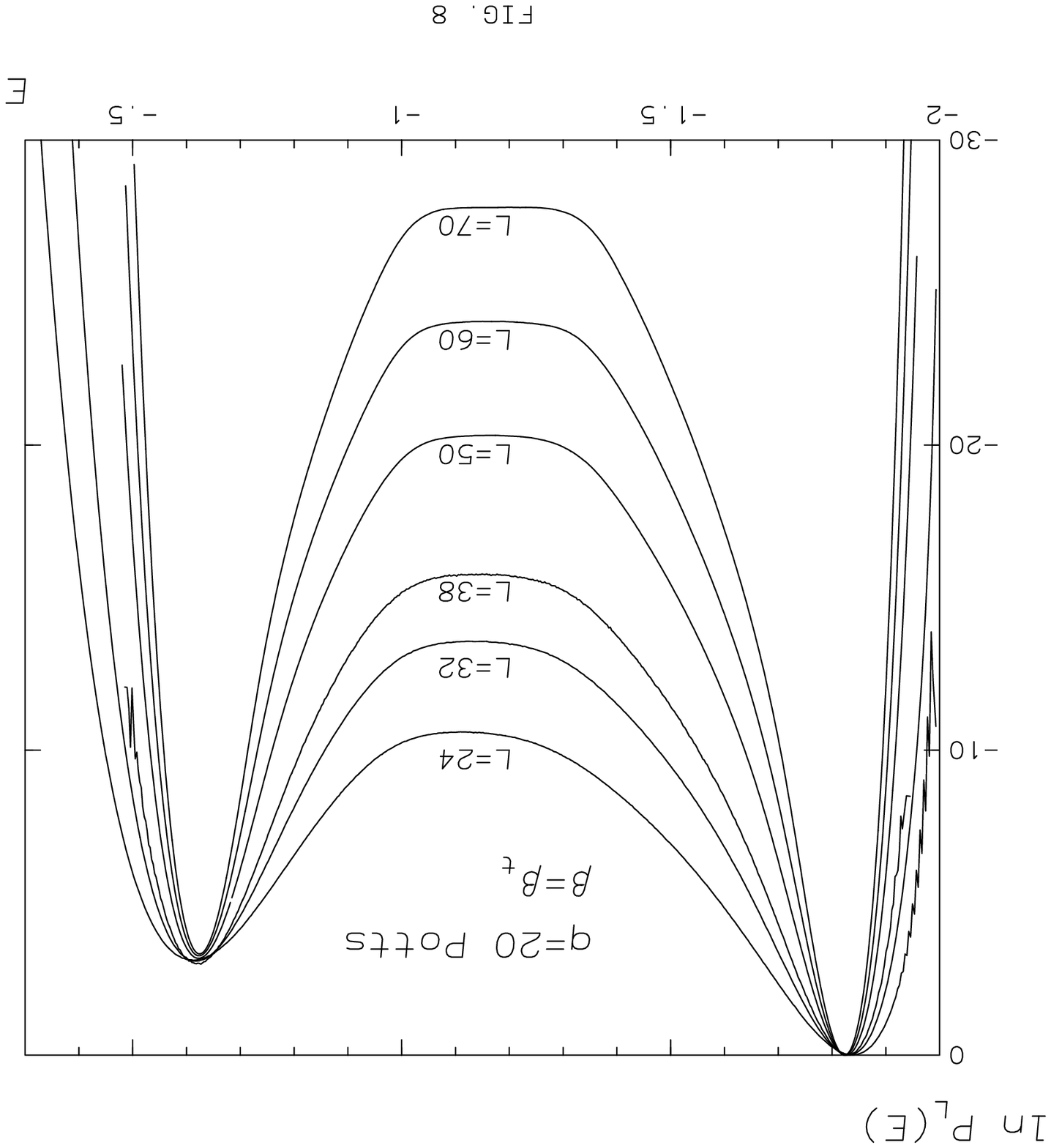}
\caption{ Plot of $\ln~P_L(\beta_t,E)$ for selected lattice sizes in the q=20
Potts model. The unfolding of a flat region with increasing
lattice size $L\gt 38$ is observed.}
\label{distri}
\end{figure}

\begin{figure} [htbp]
\vfill\penalty -5000\vglue 14cm
\includegraphics{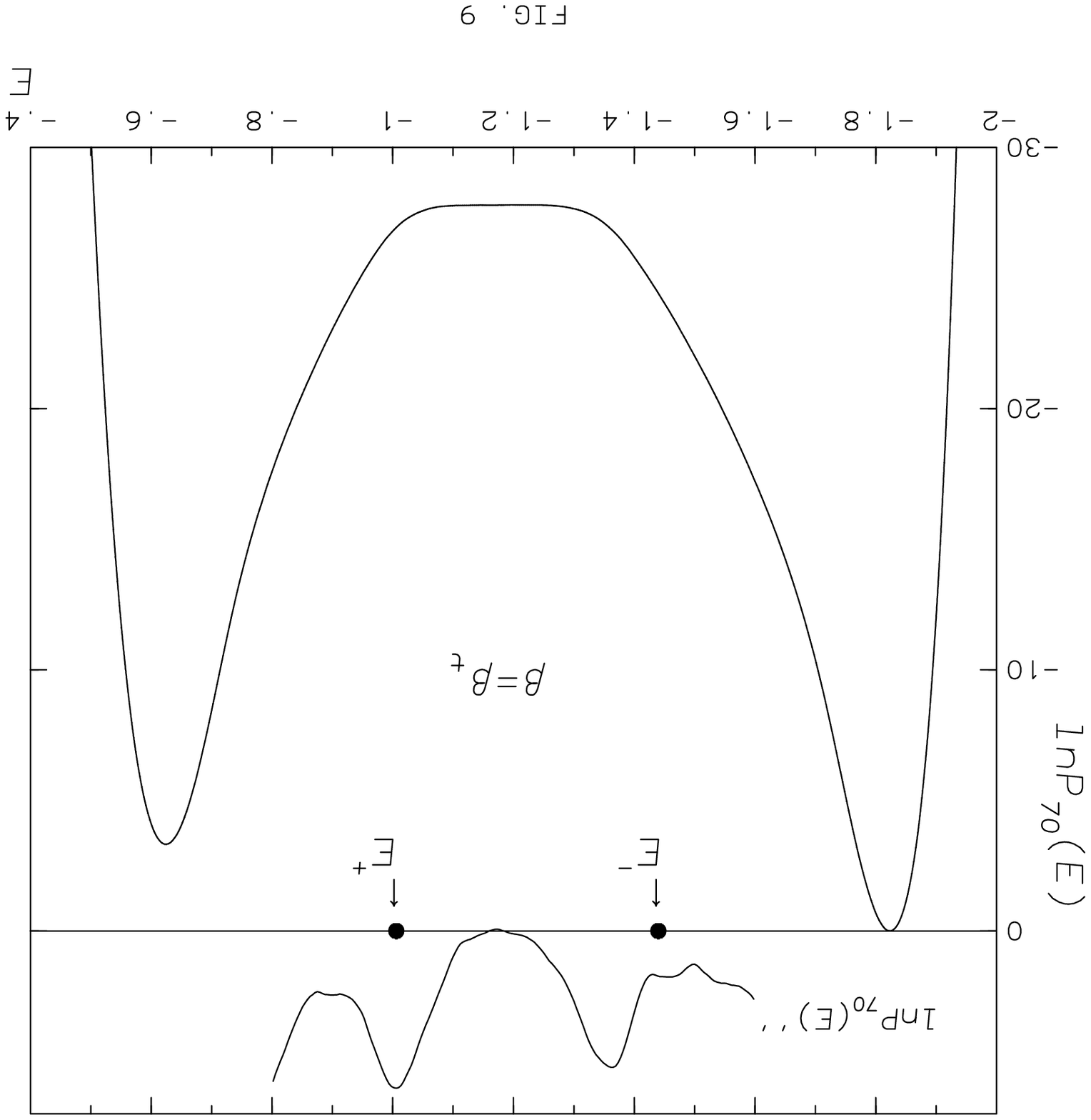}
\caption{Plot of $\ln~P_L(\beta_t,E)$ on a $70^2$ lattice. The
statistical errors cannot be resolved.
In the upper half of the plot we display a numerical estimate
of the curvature of $\ln~P_L(\beta_t,E)$. The transition from the two planar
interface to the single bubble region is indicated by two peaks.
Their position is compared with the prediction for
spherical bubbles
.}
\label{70dis}
\end{figure}

In figure \ref{distri} we display a selection of our data for the
probability distribution function $P_L(\beta_t,E)$ for the q=20 Potts model.
An analogous figure for the q=10 Potts model can be found in \cite{BN}.
We observe the unfolding of a flat region of the probability distribution
function on
our largest lattices $L > 38$, therefore giving
direct numerical evidence for the dominance of two planar interface
configurations.  On the $50^2$, $60^2$ and $70^2$ lattices we have tried to
estimate the limits of the energy range where this occurs.
For this purpose we have numerically estimated the curvature of the
probability distribution function. We assume that on finite
lattices the locations of largest curvature determine the limits.
An example of this procedure is displayed in figure \ref{70dis}.
The obtained values, as collected in Table 5, are rather close to the
values $E^{\pm}$ of the spherical bubble  model.

\begin{figure} [htbp]
\vfill\penalty -5000\vglue 14cm
\includegraphics{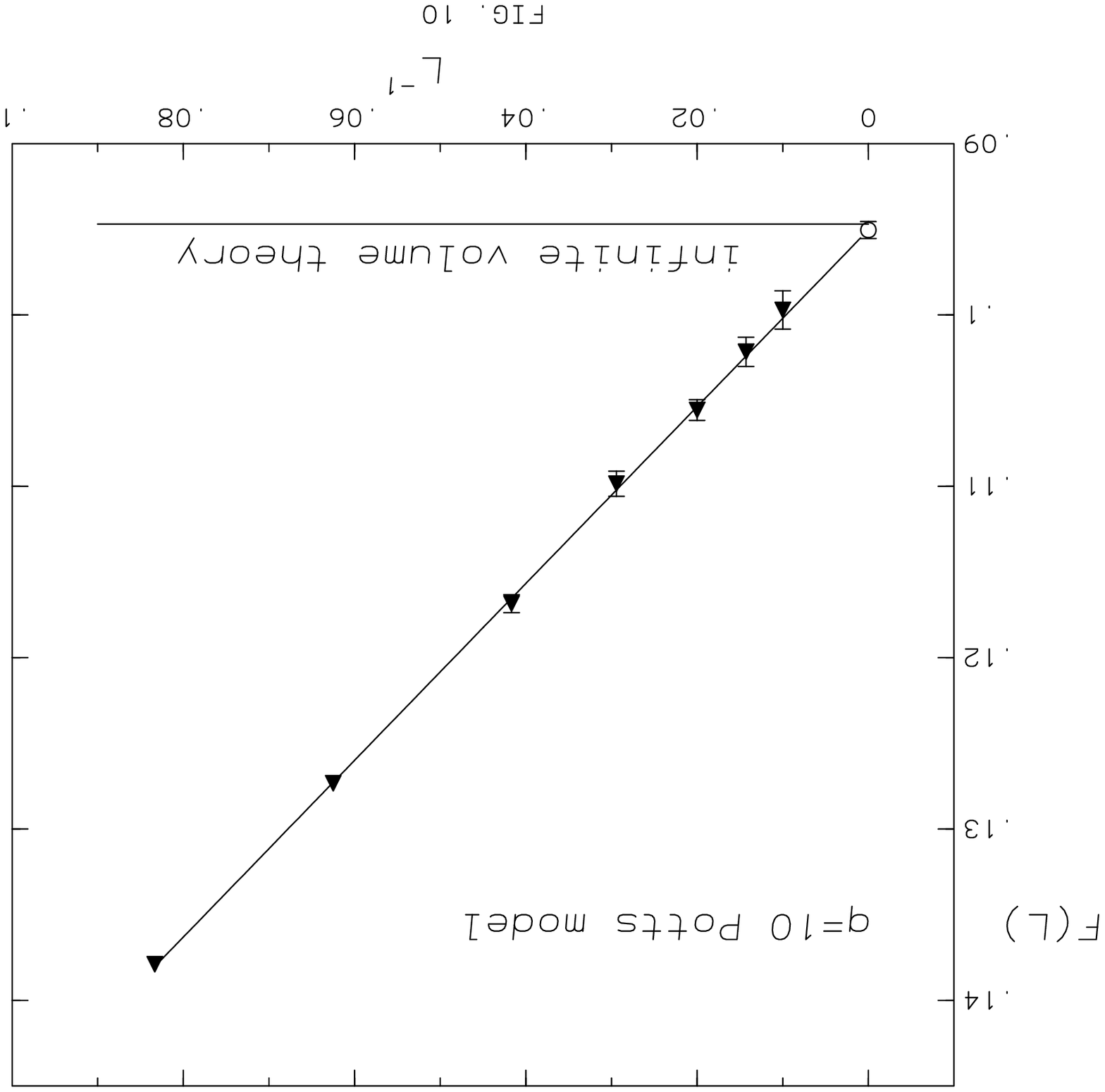}
\caption{Plot of the finite volume estimator  $F(L)$ for the q=10
Potts model as function of $1/L$. The horizontal line
corresponds to the theoretical value, and a fit of the form
$F(L)=F_{o.d.} +a_1/L$ is displayed.
}
\label{fs10}
\end{figure}

\begin{figure} [htbp]
\vfill\penalty -5000\vglue 14cm
\includegraphics{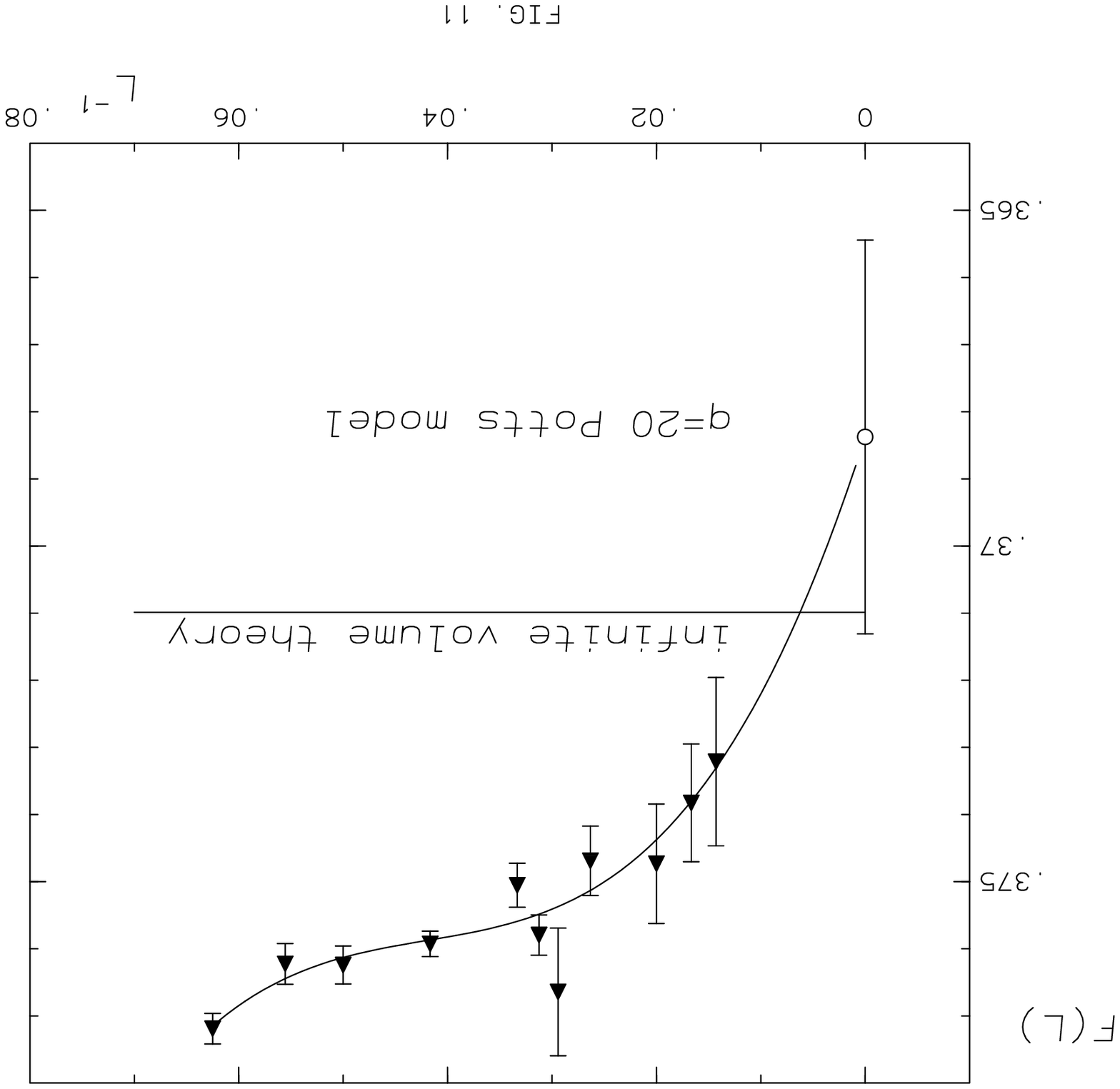}
\caption{Plot of the finite volume estimator  $F(L)$ for the q=20
Potts model as function of $1/L$. The horizontal line
corresponds to the theoretical value, and a fit of the form
$F(L)=F_{o.d.} +a_1/L+a_2/L^2+a_3/L^3$ is displayed.
}
\label{fs20}
\end{figure}

  Tables 3 and 4 also contain our results for the finite
volume estimators of the interface free energy $F(L)$.
They have been plotted in figures \ref{fs10} and \ref{fs20} as a function
of the variable $1/L$. The horizontal line in these
figures always denotes the known analytical value \cite{BuffWall}.
It can be seen that in case of the q=10 model the
 finite size behavior of $F(L)$ is very well described, in the
whole $L$ range, by a
correction proportional to $1/L$ ({\it i.e.} $p=1$) as
predicted by the capillary wave approximation. However in the
q=20 Potts model we find a more complicated behavior.
We cannot draw a
conclusion on the actual form of the functions $F(L)$, {\it i.e.} it is
undecided whether logarithmic corrections are present or not.

We have performed several fits with the form eq.(\ref{fitform}).
We have successively freed higher order
corrections in $1/L$ and also varied the interval of lattices
sizes. The results of this fits together
with their $\chi^2 / d.o.f.$ values are collected in Tables 6 and 7. All
fits of Table 6 and most fits of Table 7  are within one
standard deviation consistent with the exact $F_{o.d.}$ result of table 1
(In the $q=10$ case, this consistency is obtained although we do not
observe the expected flat part in $P_L(\beta_t,E)$, in the
$q=20$ case this consistency requires either to omit
small lattice data, or to include $1/L^2$ corrections).

As our final result we quote
\begin{equation}
F_{o.d}=0.0950(5)  ~{\rm for}~ q=10
\end{equation}
and
\begin{equation}
F_{o.d}=0.3714(13) ~{\rm for}~ q=20.
\end{equation}

\section{Conclusion}

We have presented a careful analysis of properties
of the probability distribution function $P_L(\beta,E)$ in the
2-d q=10 and q=20 Potts model at the infinite volume
transition point.

$\bullet$ On our largest lattices the ratio of
the heights of the two maxima of $P_L(\beta_t,E)$
follows the prediction of the Gaussian model.
The ratio of the ordered to the disordered partition function is equal to q.
The finite volume corrections to the locations
of the maxima of the energy follow a $1/L^d$ behavior
with a slope as predicted by the large $q$ expansion.

$\bullet$ The Gaussian width of the peaks of $P_L(\beta_t,E)$ approaches for
large volumes the prediction of the Gaussian model with specific heat
parameters $C_o$ and $C_d$, which
in the case of the q=20 Potts model are very well consistent
with the large $q$ expansion and a determination using finite size
scaling. In the $q=10$
case lattices much larger then $100^2$ would be required in order to
see the asymptotic behavior.

$\bullet$ The finite volume interface free energy estimators allow in
both cases a natural extrapolation of the data to the recently calculated
exact infinite volume values. Our results are consistent
with those values.

$\bullet$ Around $E_L^{min}$ we observe the unfolding of a flat region in
$P_L(\beta_t,E)$. It is an interesting recent observation that this
effect can be enhanced by using lattices which are elongated in one
direction \cite{GL}.
\hfill\break

{\bf Acknowledgements:} This research was partially supported by the
U.S. Department of Energy under Contracts DE-FG05-87ER40319 and
DE-FC05-85ER2500. We acknowledge conversations with Andr\'e Morel.

\begin{table}[*]
\centering
\begin{tabular}{||c|c|c||}                    \hline
 q & $F_{o.d.}$  & MC estimate             \\ \hline
  7  &  0.020792 & 0.0241 (10)  \cite{JBK} \\ \hline
 10  &  0.094701 & 0.0978 (08) ~\cite{BN}  \\ \hline
 20  &  0.370988 & $-$                     \\ \hline
\end{tabular}
\caption{{\em The exact interface free energy  for the 2-d q=7, 10
and 20 Potts models, together with previous MC estimates.
}}
\end{table}

\begin{table}[*]
\centering
\begin{tabular}{||c|c||}                    \hline
 $L$ & Msweeps \\ \hline
 50  &  26       \\ \hline
 60  &  22       \\ \hline
 70  &  10       \\ \hline
\end{tabular}
\caption{{\em Additional statistics on $50^2,60^2$, and $70^2$ lattices
in the q=20 Potts models in units of million sweeps.
}}
\end{table}

\begin{table}[*]
\centering
\begin{tabular}{||c|c|c|c|c|c||}                    \hline
 L  & $C_o$     & $C_d-C_o$& $E^{max,d}$ & $R_{10}$              & $F(L)$ \\
    & $C_d$     &          & $E^{max,o}$ & $Z_o/Z_d$             &  \\ \hline
 12 & ~4.81(06) & 0.97(12) & ~-.8813(08) &  10.70(10) & 0.1379(04) \\
    & ~5.77(09) &          & -1.7666(05) &  11.00(09) &       \\ \hline
 16 & ~5.33(04) & 1.09(08) & ~-.9012(04) &  10.25(15) & 0.1273(04) \\
    & ~6.45(06) &          & -1.7399(04) &  ~9.76(13) &       \\ \hline
 24 & ~7.06(05) & 0.77(09) & ~-.9243(06) &  ~9.98(28) & 0.1168(05) \\
    & ~7.84(06) &          & -1.7112(05) &  ~9.46(25) &       \\ \hline
 34 & ~8.64(04) & 0.82(06) & ~-.9391(02) &  10.39(57) & 0.1098(07) \\
    & ~9.43(05) &          & -1.6954(02) &  ~9.88(51) &       \\ \hline
 50 & 10.66(09) & 0.37(16) & ~-.9504(03) &  10.56(74) & 0.1056(06) \\
    & 11.14(11) &          & -1.6835(03) &  10.33(69) &        \\ \hline
 70 & 12.60(14) & 0.48(28) & ~-.9568(05) &  10.2 (12) & 0.1022(09) \\
    & 13.08(23) &          & -1.6760(04) &  ~9.7 (11) &        \\ \hline
100 & 13.99(20) & 0.54(45) & ~-.9612(05) &  ~9.2 (17) & 0.0997(11) \\
    & 14.71(35) &          & -1.6712(02) &  ~9.1 (16) &        \\ \hline
$\infty$ & 18.06(4) & 0.44763 & ~-.96820  & 10.12  & 0.094701   \\
    & 18.51(4)     &          & -1.66425  & 10.0  &            \\ \hline
\end{tabular}
\caption{{\em Finite volume results for the q=10 Potts model
at $\beta_t$, together with theoretical expectations in the
infinite volume limit.
}}
\end{table}

\begin{table}[*]
\centering
\begin{tabular}{||c|c|c|c|c|c||}                    \hline
 L  & $C_o$       & $C_d-C_o$  & $E^{max,d}$ & $R_{20}             $ & $F(L)$
\\
    & $C_d$       &            & $E^{max,o}$ & $Z_o/Z_d$             &  \\
\hline
16  & 3.473(10) & 0.742(14) & ~-.60079(09) &  22.40(16) & 0.3772(02) \\
    & 4.215(08) &           & -1.85270(09) &  20.08(13) &         \\ \hline
18  & 3.486(09) & 0.912(16) & ~-.60482(13) &  22.71(28) & 0.3762(03) \\
    & 4.403(12) &           & -1.84726(10) &  20.41(25) &         \\ \hline
20  & 3.768(32) & 0.789(36) & ~-.60796(09) &  22.19(26) & 0.3762(03)  \\
    & 4.556(10) &           & -1.84307(10) &  20.16(23) &         \\ \hline
24  & 3.895(08) & 0.909(12) & ~-.61236(07) &  22.62(24) & 0.3759(02) \\
    & 4.808(08) &           & -1.83765(06) &  20.61(22) &         \\ \hline
30  & 4.196(12) & 0.908(16) & ~-.61653(09) &  21.44(39) & 0.3751(03) \\
    & 5.109(12) &           & -1.83265(07) &  19.66(36) &         \\ \hline
32  & 4.270(13) & 0.942(17) & ~-.61761(09) &  21.66(65) & 0.3758(03) \\
    & 5.213(14) &           & -1.83149(07) &  19.85(59) &         \\ \hline
34  & 4.370(26) & 0.831(35) & ~-.61856(14) &  22.87(46) & 0.3766(10) \\
    & 5.204(36) &           & -1.83048(11) &  20.98(36) &         \\ \hline
38  & 4.495(13) & 0.863(24) & ~-.61980(08) &  20.84(68) & 0.3747(05) \\
    & 5.356(18) &           & -1.82877(06) &  19.24(63) &         \\ \hline
50  & 4.739(20) & 0.864(36) & ~-.62236(14) &  23.9 (24) & 0.3747(09) \\
    & 5.603(26) &           & -1.82572(09) &  22.1 (22) &         \\ \hline
60  & 4.925(19) & 0.768(30) & ~-.62334(14) &  26.0 (41) & 0.3738(09) \\
    & 5.693(30) &           & -1.82410(10) &  24.3 (38) &         \\ \hline
70  & 5.004(28) & 0.821(42) & ~-.62430(13) &  27.0 (45) & 0.3732(13) \\
    & 5.825(31) &           & -1.82334(10) &  25.1 (42) &         \\ \hline
$\infty$ &  5.362(3) & 0.77139   & ~-.62653    & 21.5       & 0.370988 \\
         &  6.133(3) &           & -1.82068    & 20.0       &           \\
\hline
\end{tabular}
\caption{{\em Finite volume results for the q=20 Potts model
at $\beta_t$, together with theoretical expectations in the
infinite volume limit.
}}
\end{table}

\begin{table}[*]
\centering
\begin{tabular}{||c|c|c||}                    \hline
 $L$    & $E_L^+$ & $E_L^-$   \\ \hline
 50     &  -1.00  &  -1.35    \\ \hline
 60     &  -1.00  &  -1.32    \\ \hline
 70     &  -1.01  &  -1.36    \\ \hline
 bubble &  -1.01  &  -1.44    \\ \hline
\end{tabular}
\caption{{\em
Prediction of the spherical bubble model (last line) together
with estimated locations on $50^2,\ 60^2$ and $70^2$ lattices
in the q=20 Potts model.
}}
\end{table}

\begin{table}[*]
\centering
\begin{tabular}{||c|c|c|c||}                      \hline
fit & $L_{min}$ & $F_{o.d.}$ & $\chi^2/d.o.f$  \\ \hline
$a_2=a_3=b=0$(*)   &  12 &   0.0950(05) &  .218  \\
                   &  16 &   0.0949(06) &  .220  \\
                   &  24 &   0.0946(08) &  .206  \\
                   &  34 &   0.0949(13) &  .257  \\
                   &  50 &   0.0938(20) &  .003  \\ \hline
$a_3=b=0$          &  12 &   0.0946(09) &  .176  \\
                   &  16 &   0.0943(13) &  .218  \\
                   &  24 &   0.0947(23) &  .307  \\
                   &  34 &   0.0924(39) &  .041  \\ \hline
$b=0$              &  12 &   0.0942(19) &  .221  \\
                   &  16 &   0.0945(33) &  .325  \\
                   &  24 &   0.0905(64) &  .103  \\ \hline
$a_2=a_3=0$        &  12 &   0.0941(15) &  .167  \\
                   &  16 &   0.0939(22) &  .217  \\
                   &  24 &   0.0945(40) &  .308  \\
                   &  34 &   0.0903(69) &  .062  \\  \hline
Theory             &     &   0.094701~~ &        \\  \hline
\end{tabular}
\caption{{\em Results of fits of the form eq.(16) to the finite
volume estimators $F(L)$ of Table 3 for the q=10 Potts model.
Column one specifies which parameters of the fit where fixed to 0,
with $b=-p+d/2$.
The fit intervals are from $L_{min}$ of column two to $L_{max}=100$.
The fit marked with (*) denotes the fit which results into our infinite
volume estimate eq.(18). It has been plotted
in figure 10.
}}
\end{table}

\begin{table}[*]
\centering
\begin{tabular}{||c|c|c|c|c||}                    \hline
fit & $L_{min}$ &  $F_{o.d.}$ & $\chi^2/d.o.f$  & consistent  \\
    &           &             &                 & with theory \\ \hline
$a_2=a_3=b=0$      &  16 &    0.3736(04) & 1.187 &   no \\
                   &  18 &    0.3739(05) & 1.184 &   no \\
                   &  20 &    0.3736(06) & 1.179 &   no \\
                   &  24 &    0.3733(07) & 1.306 &   no \\
                   &  30 &    0.3726(12) & 1.470 &   no \\
(*)                &  32 &    0.3714(13) &  .590 &  yes \\
                   &  34 &    0.3713(17) &  .782 &  yes \\
                   &  38 &    0.3724(19) &  .265 &  yes \\
                   &  50 &    0.3694(45) &  .001 &  yes \\ \hline
$a_3=b=0$          &  16 &    0.3733(12) & 1.330 &   no \\
                   &  18 &    0.3717(15) &  .985 &  yes \\
                   &  20 &    0.3716(17) & 1.149 &  yes \\
                   &  24 &    0.3709(23) & 1.329 &  yes \\
                   &  30 &    0.3663(46) & 1.335 &   no \\
                   &  32 &    0.3726(59) &  .773 &  yes \\
                   &  34 &    0.3766(94) & 1.009 &  yes \\
                   &  38 &    0.362~(13) &  .028 &  yes \\ \hline
$b=0$(**)          &  16 &    0.3684(29) & 1.033 &  yes \\
                   &  18 &    0.3701(38) & 1.116 &  yes \\
                   &  20 &    0.3687(46) & 1.287 &  yes \\
                   &  24 &    0.363~(10) & 1.528 &  yes \\
                   &  30 &    0.390~(21) & 1.368 &  yes \\
                   &  32 &    0.358~(31) & 1.057 &  yes \\ \hline
$a_2=a_3=0$        &  16 &   0.3727(20)  & 1.312 &  yes \\
                   &  18 &   0.3698(25)  &  .962 &  yes \\
                   &  20 &   0.3699(30)  & 1.121 &  yes \\
                   &  24 &   0.3688(40)  & 1.312 &  yes \\
                   &  30 &   0.3608(88)  & 1.383 &   no \\
                   &  32 &   0.373~(11)  &  .777 &  yes \\
                   &  34 &   0.379~(18)  & 1.064 &  yes \\
                   &  38 &   0.354~(25)  &  .037 &  yes \\ \hline
Theory             &     &   0.370988~~  &       &      \\ \hline
\end{tabular}
\caption{{\em Results of fits of the form eq.(16) to the finite
volume estimators $F(L)$ of Table 4 for the q=20 Potts model.
Column one specifies which parameters of the fit where fixed to 0,
with $b=-p+d/2$.
The fit intervals are from $L_{min}$ of column two to $L_{max}=70$.
The last column specifies whether the exact $F_{o.d.}$ result is within
one standard deviation of the fitted value.
The fit marked with (*) denotes the fit which results into our infinite
volume estimate eq.(19), the one marked with (**) has been plotted
in figure 11.
}}
\end{table}
\hfill\break\vfill\eject
\clearpage


%

\hfill\break\vfill\eject

\end{document}